\documentclass[sigconf]{acmart}

\usepackage{amsmath}
\usepackage[ruled,linesnumbered]{algorithm2e}
\usepackage{soul}

\usepackage{comment}
\usepackage{textcomp}
\usepackage[keeplastbox]{flushend}

\usepackage{comment}

\newcommand{\Remove}[1]{{}}

\newcommand{\squishlist}{
   \begin{list}{$\bullet$}
    { \setlength{\itemsep}{0pt}      \setlength{\parsep}{0pt}
      \setlength{\topsep}{3pt}       \setlength{\partopsep}{0pt}
      \setlength{\listparindent}{-2pt}
      \setlength{\itemindent}{-5pt}
      \setlength{\leftmargin}{1em} \setlength{\labelwidth}{0em}
      \setlength{\labelsep}{0.5em} } }

\newcommand{\squishend}{
    \end{list}  }
    
\usepackage{multirow}  
\usepackage{comment}
\usepackage[ruled,linesnumbered]{algorithm2e}
\usepackage{listings}
\usepackage{xcolor}
\usepackage{footnote}
\usepackage{threeparttable}
\usepackage{paralist}
\usepackage{booktabs}

\usepackage{xcolor}
\usepackage[normalem]{ulem} 

\usepackage{tikz}
\usetikzlibrary{calc}

\newcommand*\blackcircled[1]{\tikz[baseline=(char.base)]{
        \node[shape=circle,fill={rgb,255:red,0;green,0;blue,0}, text=white, font=\small, inner sep=0.6pt] (char) {#1};}}

\usepackage{multirow}

\usepackage{framed}

\usepackage{pifont}

\usepackage[ruled,linesnumbered]{algorithm2e}
\usepackage{xcolor}
\usepackage{booktabs}
\usepackage{graphicx}  

\usepackage{wrapfig}

\usepackage{fancyhdr}
\renewcommand\footnotetextcopyrightpermission[1]{}

\usepackage{ifthen}
\usepackage{amsmath}

\usepackage[utf8]{inputenc}
\usepackage[ruled,linesnumbered]{algorithm2e}

\usepackage{ulem}

\usepackage{subfigure}

\AtBeginDocument{%
  \providecommand\BibTeX{{%
    \normalfont B\kern-0.5em{\scshape i\kern-0.25em b}\kern-0.8em\TeX}}}

\setcopyright{acmcopyright}
\copyrightyear{2022}
\acmYear{2022}
\acmDOI{10.1145/1122445.1122456}

\acmConference[Woodstock '18]{Woodstock '18: ACM Symposium on Neural
  Gaze Detection}{June 03--05, 2018}{Woodstock, NY}
\acmBooktitle{Woodstock '18: ACM Symposium on Neural Gaze Detection,
  June 03--05, 2018, Woodstock, NY}
\acmPrice{15.00}
\acmISBN{978-1-4503-XXXX-X/18/06}

\begin{document}

\title{Alleviating Datapath Conflicts and Design \\Centralization in Graph Analytics Acceleration}



\author{Haiyang Lin$^{1,2}$, Mingyu Yan$^{1,2,*}$,Duo Wang$^{1,2}$, Mo Zou$^{1,2}$, Fengbin Tu$^{3}$, Xiaochun Ye$^{1,2}$, \\Dongrui Fan$^{1,2}$, Yuan Xie$^{3}$}

\affiliation{
    \institution{$^1$State Key Laboratory of Computer Architecture, Institute of Computing Technology, Chinese Academy of Sciences}~\state{Beijing}~\country{China}}
\affiliation{
    \institution{$^2$University of Chinese Academy of Sciences} \state{Beijing}
    \country{China}}
\affiliation{
    \institution{$^3$University of California at Santa Barbara}
    \state{CA}
    \country{USA}}



\begin{abstract}
%


Previous graph analytics accelerators have achieved great improvement on throughput by alleviating irregular off-chip memory accesses. 
However, on-chip side datapath conflicts and design centralization have become the critical issues hindering further throughput improvement.
In this paper, a general solution, \uline{M}ultiple-stage \uline{D}ecentralized \uline{P}ropagation network (MDP-network), is proposed to address these issues, inspired by the key idea of trading latency for throughput. 
Besides, a novel \uline{Hi}gh throughput \uline{Graph} analytics accelerator, HiGraph, is proposed by deploying MDP-network to address each issue in practice. 
The experiment shows that compared with state-of-the-art accelerator, HiGraph achieves up to 2.2$\times$ speedup (1.5$\times$ on average) as well as better scalability.

\end{abstract}

\keywords{graph analytics, datapath conflicts, design centralization, acceleration, domain-specific architecture}

\maketitle

\thispagestyle{fancy}         
\fancyhead{}                     
\lhead{}          
\chead{\ifthenelse{\value{page}=1}{To Appear in 59th Design Automation Conference (DAC 2022)}{}}
\rhead{}
\lfoot{}
\cfoot{}  
\rfoot{}
\renewcommand{\headrulewidth}{0pt}      
\renewcommand{\footrulewidth}{0pt}

\section{Introduction}

Graphs exhibit powerful representation capacity for real-world data in a broad range of scenarios, which is accompanied by many graph analytics applications, such as placement \cite{EDA:placement1}, circuit partitioning \cite{EDA:cp1}, and technology mapping \cite{EDA:tm1} in EDA flow.
Not surprisingly, the demand for high throughput in the execution of graph analytics workloads is ever-growing as the scale of graph data increases.




Recently, plenty of graph analytics accelerators have been well designed to improve throughput by alleviating the critical performance bottleneck, i.e., irregular accesses to off-chip memory. 
For example, Graphicionado \cite{Graphicionado} and GraphDynS \cite{GraphDynS} leverage a large-capacity on-chip memory to buffer all vertices' property data on the chip, significantly alleviating irregular off-chip accesses. Even in the case of large-scale graphs, graph slicing can be used to partition the graph into a set of small slices to perform the processing using limited capacity of on-chip memory \cite{Graphicionado}.
%
%

Unfortunately, on-chip side datapath conflicts and design centralization are becoming increasingly significant bottlenecks on the throughput improvement. 
To exploit the high-degree parallelism in graph analytics workloads, most of the accelerators are in favor of adopting multiple parallel execution channels. 
However, due to the irregular connection pattern across vertices, interaction across execution channels is inevitable, which brings two following inefficiencies to previous designs. 
The first is \textbf{Datapath Conflicts}, \emph{which means multiple datapaths that process different vertices compete for the same accessible resource or same dataflow channel, causing serious stall in datapaths.} As a result, the overall performance will significantly degrade due to the severe blocking of execution channels.
%
The second is \textbf{Design Centralization}, \emph{which means the implementation of design becomes extremely difficult with the increasing amount of execution channels due to the overintensive interaction across total execution channels, causing frequency decline.}
For example, on-chip crossbar is a prevalent solution to direct the dataflow between different execution channels \cite{Graphicionado, GraphDynS}.
However, it suffers from not only the frequency decline which hinders the pursuit of high throughput, but also a dramatic increase in area and power consumption, when channel number increases \cite{Crossbar_Model}. 


In this work, we observe that the execution channel in the state-of-the-art accelerator is highly pipelined~\cite{Graphicionado,GraphDynS}, which reveals that increasing the traversal latency of a single edge does not pose significant impact on overall performance. 
Therefore, inspired by the key idea of trading latency for throughput, we propose a general solution, \uline{M}ultiple-stage \uline{D}ecentralized \uline{P}ropagation network (\textbf{MDP-network}).
MDP-network decentralizes the intensive interactions across execution channels to multiple stages and buffers data in each stage.
Data in MDP-network is propagated deterministically to next stage until reaching their destinations. 
On one hand, the multiple-stage and deterministic propagation alleviate datapath conflicts.

On the other hand, the inefficiency of design centralization is avoided since the number of interactive execution channels in each stage is limited to a small number. 
To facilitate adoption, we provide an open-source automatic generator of MDP-network\footnote{https://github.com/OpenSource88/MDP-network.git}.
Finally, a novel \uline{Hi}gh throughput \uline{Graph} analytics accelerator, \textbf{HiGraph}, is proposed by deploying MDP-network to tackle data conflicts and design centralization in practice.

The main contributions of this paper are as follows:
\squishlist
\item We identify the inefficiencies including datapath conflicts and design centralization in graph analytics acceleration.
\item We propose MDP-network, Multiple-stage Decentralized Propagation network, to alleviate datapath conflicts and design centralization. Besides, an automation tool is developed to generate MDP-network and open source to facilitate its deployment.
\item We propose HiGraph, a novel high throughput graph analytics accelerator coupled with MDP-network, and implement it in RTL. The experimental results show that compared to the state-of-the-art design, HiGraph achieves up to 2.2$\times$ speedup (1.5$\times$ on average) as well as better scalability.
\squishend

\section{Background and Motivation}



\subsection{CSR Format and VCPM}

\begin{figure}[hbtp]
    \vspace{-10pt}
    \centering
    \includegraphics[page=1, width=0.4\textwidth]{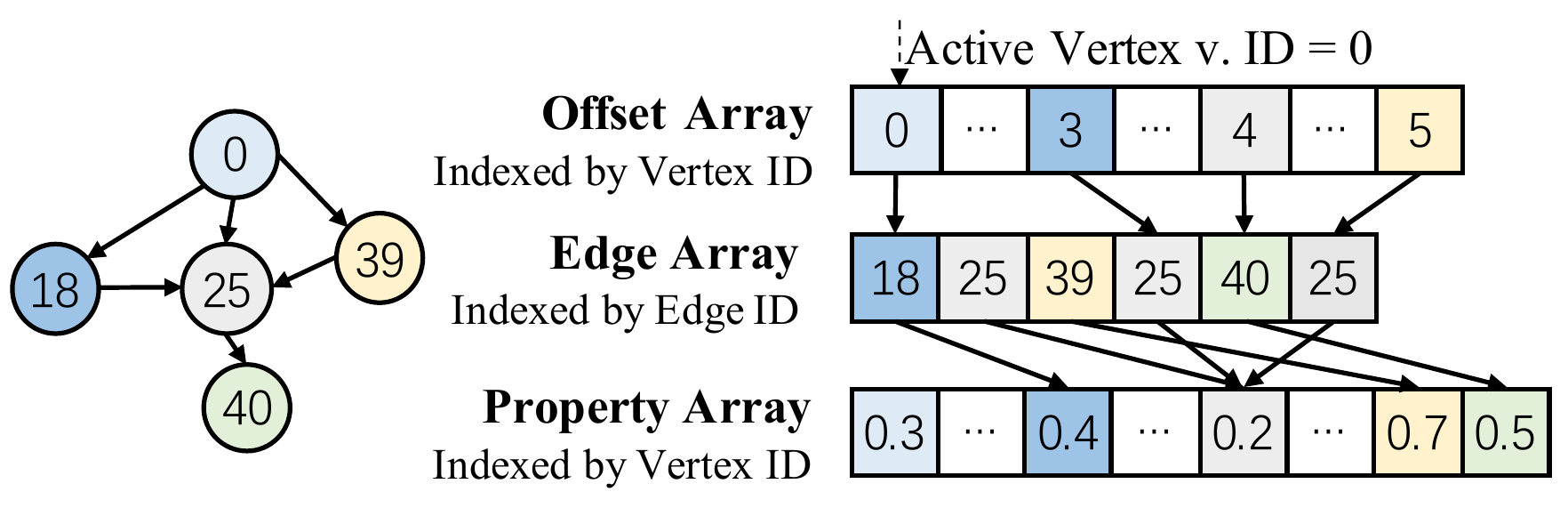}
    \vspace{-10pt}
    \caption{An example graph in CSR format.}
    \label{fig:000CSR}
    \vspace{-10pt}
\end{figure}

Compressed Sparse Row (CSR) format is a widely used storage-efficient technique to represent graph structures in software frameworks \cite{soft:GraphX, soft:Ligra} and accelerators \cite{Graphicionado, GraphDynS}. Fig. \ref{fig:000CSR} illustrates that three data arrays, \textit{Offset}, \textit{Edge}, and \textit{Property}, are used to encode a graph. Each Offset entry stores the position of its first neighbor in the Edge Array. The Edge Array maintains destination vertex ID and weight for each outgoing edge. The Property Array holds current property value for each vertex.

\begin{figure*}[!htbp] 
    \centering
    \begin{minipage}{0.30\textwidth}
    \vspace{-2pt}
    \centering
    \includegraphics[width=\textwidth]{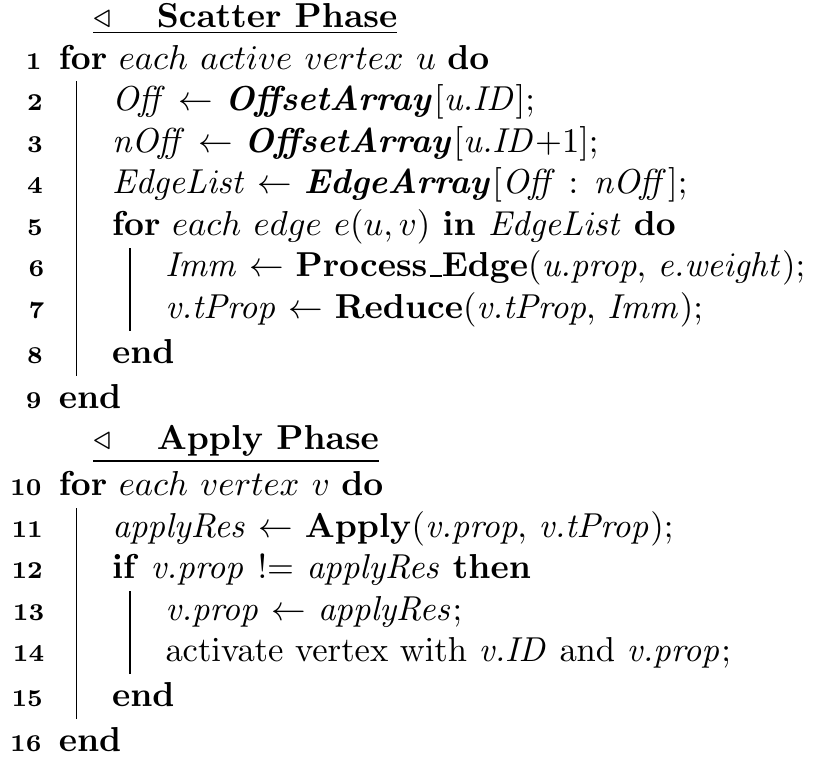}
    \vspace{-22pt}
    \caption{Pseudocode of VCPM.}
    \label{fig:001vcpm}
    \end{minipage}\hfill
    \begin{minipage}{0.69\textwidth}
    \centering
    \includegraphics[page=2, width=\textwidth]{real_graph.pdf}
    \vspace{-22pt}
    \caption{The structure of parallel execution channels in VCPM-based graph analytics accelerator (top) and three types of datapath conflicts (bottom).}
    \label{fig:001conflicts}
    \end{minipage}\hfill
    \vspace{-5pt}
\end{figure*}

Existing graph analytics software frameworks \cite{soft:Pregel_vcpm, soft:MapGraph_vcpm} and accelerators \cite{Graphicionado, GraphDynS} usually employ \textit{Vertex-Centric Programming Model} (VCPM) to accomplish iterative graph algorithms.
VCPM consists of \textit{scatter} and \textit{apply} phases, as shown in Fig. \ref{fig:001vcpm}. In the scatter phase, each active vertex first aggregates the effect of its property and edge weight via a user-defined function Process$\_$Edge(~), then broadcasts the accumulated influence to update its outgoing neighbors in an additional tProperty Array using user-defined function Reduce(~). 
In the apply phase, data in the tProperty Array is synchronized to the Property Array using user-defined function Apply(~).
VCPM updates the Property Array iteratively until all vertices are inactive.

\subsection{Inefficiencies of Previous Designs} \label{sec:inefficiency}

By adopting multiple parallel execution channels, as shown in Fig. \ref{fig:001conflicts}, previous VCPM-based graph analytics accelerators have achieved great throughput improvement. 
However, due to irregular connection pattern across vertices, interaction across execution channels is inevitable, bringing two following inefficiencies in previous designs.

\textbf{Datapath Conflicts:} multiple datapaths that process different vertices compete for the same accessible resource or same dataflow channel, causing serious datapath stall.
Fig. \ref{fig:001conflicts} demonstrates three types of datapath conflicts existing in graph analytics accelerators with multiple parallel execution channels.
Note that to meet the requirement of data-access throughput in such a design, the buffer for each data array is divided into several parts and organized in the fashion of interleaving. 

The first datapath conflict occurs in \blackcircled{1} \emph{Offset Array Access}, where two consecutive buffer parts are accessed concurrently to obtain start and end positions in the Edge Array of one active vertex. 
The second datapath conflict happens in \blackcircled{2} \emph{Edge Array Access}, where a list of edges is accessed simultaneously from multiple buffer parts. 
The third datapath conflict occurs in \blackcircled{3} \emph{Dataflow Propagation}, where the dataflow is directed according to the destination vertex ID of each edge. 
Datapath conflicts arise when multiple datapaths require the same accessible resource or same dataflow channel simultaneously. 
As a result, the execution channels failing in arbitration will be blocked, which inevitably degrades overall performance.

\begin{figure}[hbtp]
    \vspace{-10pt}
    \centering
    \includegraphics[width=0.32\textwidth]{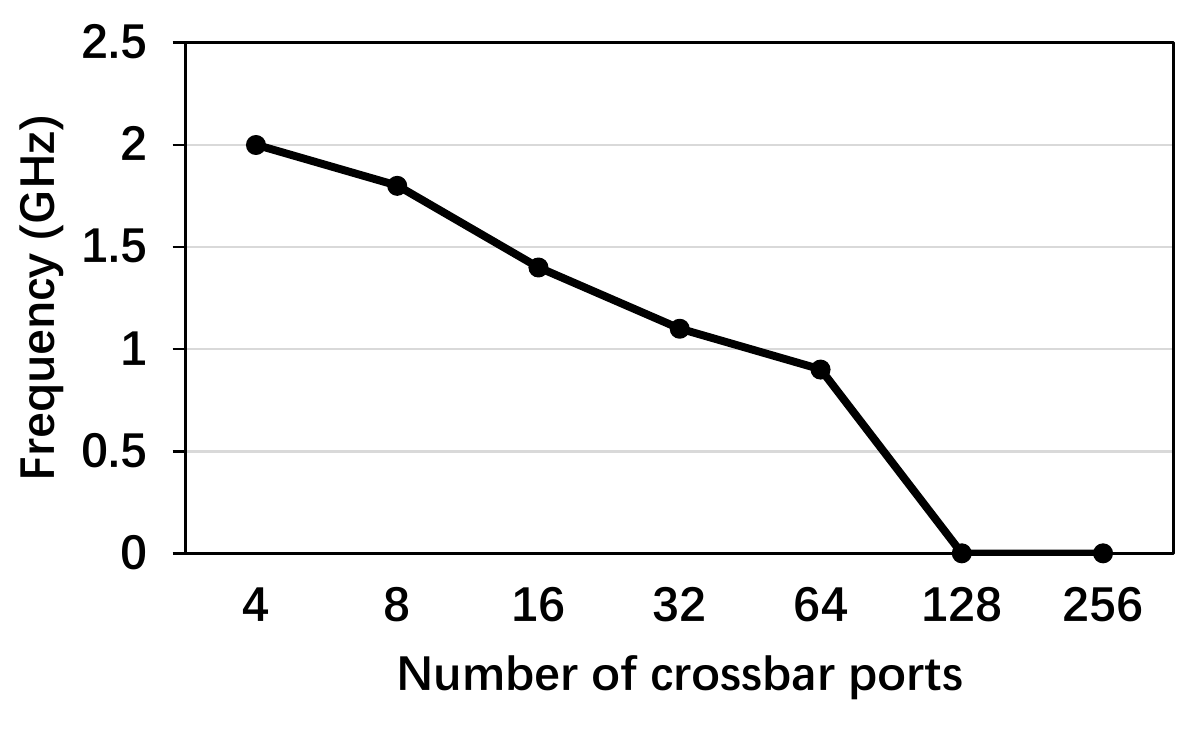}
    \vspace{-10pt}
    \caption{Frequency versus number of crossbar ports.}
    \label{fig:002freq_cb}
    \vspace{-10pt}
\end{figure}

\textbf{Design Centralization:} the implementation of design becomes extremely difficult with the increasing amount of execution channels due to the overintensive interaction across total execution channels, causing frequency decline.
In previous graph analytics accelerators, arbitration solution like crossbar is prevalently used to deal with interaction of multiple channels \cite{Graphicionado, GraphDynS}.
Fig. \ref{fig:002freq_cb} demonstrates that the frequency declines sharply with the increasing number of crossbar ports, limiting the throughput improvement. The frequency is determined by the critical path time which is derived from the synthesis result of Synopsys Design Compiler.


\textbf{Opportunity:} we observe that the execution channel in the state-of-the-art accelerator is highly pipelined, which reveals that increasing the traversal latency of a single edge has marginal impact on overall performance.
It inspires us that we can trade latency for throughput. Thus, it is practical to insert additional stages in the datapath to gradually guide the data to the destination execution channel, which alleviates datapath conflicts. Besides, we can alleviate design centralization by limiting the number of interactive channels in each stage. 
Based on the above insights, we propose a general solution, MDP-network, to alleviate datapath conflicts and design centralization.


\section{Multiple-stage Decentralized Propagation Network}

\begin{figure*}[hbtp]
    \vspace{-10pt}
    \centering
    \includegraphics[page=3, width=0.96\textwidth]{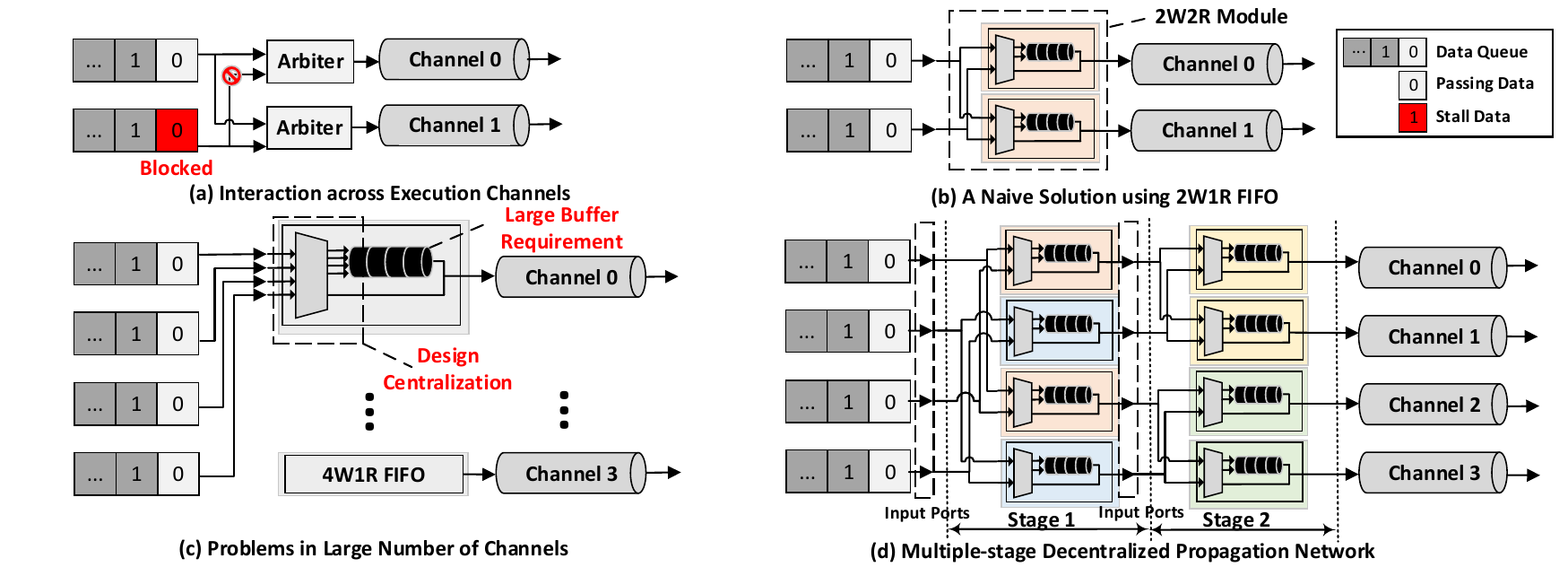}
    \vspace{-10pt}
    \caption{Design theory of MDP-network.}
    \label{fig:003idea}
    \vspace{-10pt}
\end{figure*}

\subsection{Design Theory}
In this subsection, we first abstract the interaction across execution channels, which brings datapath conflicts and design centralization. Then, we introduce the design theory of MDP-network from a naive solution.

Fig. \ref{fig:003idea} (a) abstracts the interaction across execution channels in Fig. \ref{fig:001conflicts}, where data from two input channels is directed to two output channels. Using arbitration, e.g., crossbar, encounters datapath conflicts. 
 
A naive solution for datapath conflicts is to use n-Write-1-Read (nW1R) First-In-First-Out queue (FIFO), which means FIFO can input n datums and output one datum in each cycle. 
As shown in Fig. \ref{fig:003idea} (b), 2W1R FIFO is used to direct data to destination channel in each cycle, therefore no input channels would stall unless the FIFO is full. 
However, it encounters two critical issues when the number of channels increases, as shown in Fig. \ref{fig:003idea} (c). 
One is large requirement of buffer capacity. For example, when the number of write ports is 32, the FIFO can accept data only when the remaining capacity is not less than 32. This causes large requirement and low utilization of buffer capacity. 
Another is design centralization, the increasing number of write ports makes FIFO implementation more difficult and even results in frequency decline. 

Inspired by the idea of trading latency for throughput, we propose a general solution, MDP-network, to address this general problem. As shown in Fig. \ref{fig:003idea} (d), MDP-network decentralizes the intensive interactions across execution channels to multiple stages and buffers data in each stage. Data in MDP-network is deterministically propagated to next stage until reaching their destinations. 


Benefiting from the multiple-stage and deterministic propagation dataflow, the head-of-line datum does not block data behind it and data is propagated to the destination channels stage by stage, resulting in less datapath conflicts and better throughput.
Moreover, since the number of interactive execution channels in each stage of MDP-network is limited to a small number, the implementation complexity is reduced.
As a result, MDP-network improves the scalability of the multiple parallel execution channels design without declining the frequency.



\begin{algorithm}
    \caption{MDP-network Generation Algorithm}
    \label{alg:dmp}
    {\small
    \KwIn{$n;$ \tcp*[f]{number of total channels}\\}
    \centerline{\quad \underline{{\textbf{Illustartes radix 2 as an example}}}}
    \centerline{\quad \underline{$\triangleleft$ \quad {\textbf{Step 1:\quad 2W2R module construction}}}}
    Use two 2W1R FIFOs to construct a 2W2R module\;
    \centerline{\quad \underline{$\triangleleft$ \quad {\textbf{Step 2:\quad Input ports connection}}}}
    \For(\tcp*[f]{stage $i$}){$i=0; i<log_2n; i++$}{ 
        \emph{pair\_list} $\leftarrow$ [~]\; 
        \emph{target\_group} $\leftarrow$ $2^i$\; 
        \emph{group\_base} $\leftarrow$ \emph{n} $/$ \emph{target\_group}\; 
        \emph{channel\_step} $\leftarrow$ \emph{group\_base} $/$ 2\; 
        \For(\tcp*[f]{group $j$}){$j=0; j<target\_group; j++$}{
        \emph{real\_base} $\leftarrow$ \emph{group\_base} $*$ \emph{j};
            
            \For(\tcp*[f]{~pair $k$}){$k=0; k<channel\_step; k++$}{
            
            \emph{channel\_1} $\leftarrow$ \emph{real\_base} $+$ \emph{k}\;
            \emph{channel\_2} $\leftarrow$ \emph{real\_base} $+$ \emph{k} $+$ \emph{channel\_step}\;
            \emph{pair\_list.append}([\emph{channel\_1}, \emph{channel\_2}])\;
            }
        }
        Input ports of each stage within the same pair are connected to one 2W2R module
        using the (\emph{$log_2n$} - \emph{i})th bit of address;\\  
    }
    }
\end{algorithm}

\subsection{Automatic Generator for MDP-network}
Algorithm \ref{alg:dmp} describes the simplified workflow of an automatic MDP-network generator. We define radix as the number of FIFO write ports and take radix 2 as an example here.
To facilitate the deployment of MDP-network, the algorithm consists of two steps: \textit{2W2R module construction} (line 1) and \textit{input ports connection} (line 2 - line 16). It is straightforward to construct a 2W2R model following the rule shown in Fig. \ref{fig:003idea} (b). 
Once the construction is finished, \textit{input ports connection} is performed to choose the corresponding 2W2R module for input ports in each MDP-network stage.

Fig. \ref{fig:003idea} (d) presents a toy example of \textit{input ports connection} for four channels. 
MDP-network uses two address bits, addr[0: 1], to specify four destination channels and constructs $log_24=2$ stages.
In the first stage, we classify all input channels into one group (\textit{target\_group} = 1) since they have the same target range, i.e., output channels 0-3. 
\textit{Channel\_step} is the difference between two input channel IDs connecting to one 2W2R module (\textit{channel\_step} = 2).
So we connect input ports \{0, 2\} and \{1, 3\} to two 2W2R modules respectively with addr[1]. Note that we draw two 2W1R modules using the same color in Fig. \ref{fig:003idea} (d) if they come from one 2W2R module.
In the second stage, \textit{target\_group} is 2 and \textit{channel\_step} is 1. Then we connect input ports \{0, 1\} from Group 1 and \{2, 3\} from Group 2 to two 2W2R modules with addr[0].
With MDP-network, data is propagated stage by stage until reaching destinations.

\begin{figure*}[!htbp]
    \vspace{-10pt}
    \centering
    \includegraphics[page=4, width=0.96\textwidth]{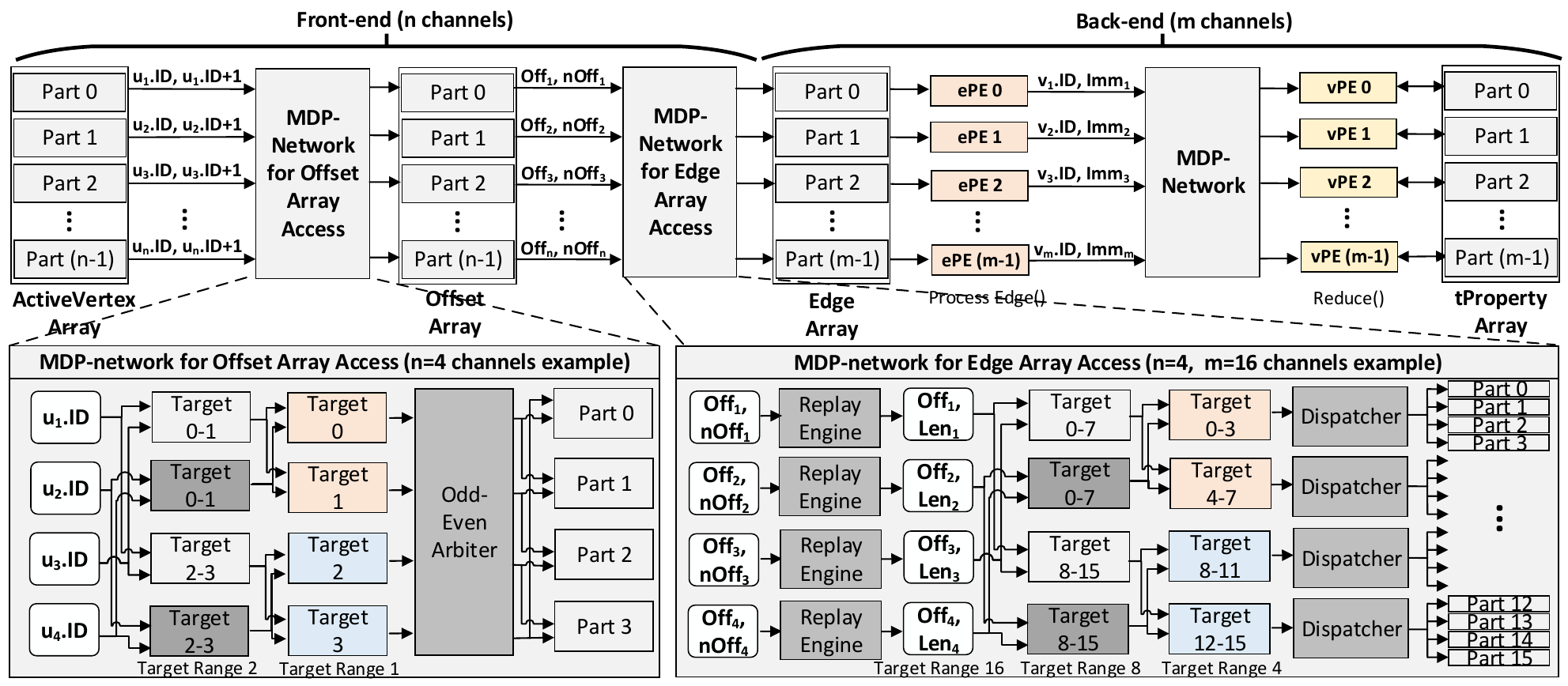}
    \vspace{-10pt}
    \caption{Architecture of HiGraph.}
    \label{fig:higraph arch}
    \vspace{-10pt}
\end{figure*}

\section{HiGraph}

In this section, we describe the architecture of HiGraph, a novel high throughput graph analytics accelerator.
HiGraph buffers all data arrays of a graph in the fashion of interleaving and adopts parallel execution channels design. 

As shown in Fig. \ref{fig:higraph arch}, the HiGraph architecture contains \textit{front-end} and \textit{back-end} parts.
The \textit{front-end} part processes the source vertex of an edge, including reading active vertex IDs from the ActiveVertex Array and obtaining corresponding edge offsets from the Offset Array.
The \textit{back-end} part executes Process\_Edge(~) and Reduce(~) after reading the Edge Array and tProperty Array, then writes back the updated value to tProperty Array.

As mentioned in Section \ref{sec:inefficiency}, there are three types of interaction across execution channels. Next, we will analyze their access patterns and address how MDP-network is deployed to ease datapath conflicts and design centralization.




\subsection{MDP-network for Offset Array Access}

Referring to Fig. \ref{fig:001conflicts} \blackcircled{1}, the access pattern in reading Offset Array is one-to-two, indicating that the algorithm reads two consecutive buffer parts indexed by \emph{u.ID} and \emph{u.ID}+1 of source vertex \emph{u}.
%

Fig. \ref{fig:higraph arch} shows that, to deal with such an access pattern, we first deploy MDP-network to guide source vertices to corresponding output channels, e.g., \emph{u.ID} 0 is directed to channel 0. 
Then it is apparently that each source vertex requires to occupy its corresponding and next read channels, e.g., vertices in channel 0 requires the read channel 0 and 1.
In this way, any source vertices in one channel will only have conflicts with those in neighbor channels.
Thus we insert an Odd-Even Arbiter to determine vertices in which channels should be issued in current cycle.

The arbiter rule is \emph{alternating priority}, which means odd and even channels alternately have higher priority to issue vertices. 
In this way, those vertices in the channel with higher priority can always be issued immediately without considering datapath conflicts and occupy several read channels.
The other vertices would be issued only when their destination read channels are not occupied or their target addresses are the same with those who have occupied the read channels.

\subsection{MDP-network for Edge Array Access}

Referring to Fig. \ref{fig:001conflicts} \blackcircled{2}, the access pattern in reading Edge Array is one-to-multiple, indicating that one \{\emph{Off}, \emph{nOff}\} requires to access multiple consecutive buffer parts to get edges.

Fig. \ref{fig:higraph arch} demonstrates the variant of MDP-network for Edge Array access pattern.
We insert Replay Engines to divide \{\emph{Off}, \emph{nOff}\} into several \{\emph{Off}, \emph{Len}\} with an appropriate length.
Then we use MDP-network to guide data to destination channels and one extra operation is required here.
While the target range is becoming smaller as the data is propagated stage by stage, correspondingly, we will split the input length into small output length to make \{\emph{Off}, \emph{Len}\} fit in small target range.
For example, when \emph{Off} 4 with \emph{Len} 9 needs to be propagated to target 0-7 and target 8-15 in stage 1, it will be split into \emph{Off} 4 with \emph{Len} 4 and \emph{Off} 8 with \emph{Len} 5.
After several stage propagation, the target range becomes smaller and more specific. Meanwhile, it can be found that there are fewer channels with the same target range. 
In other words, through the propagation of MDP-network, the competition for subsequent datapaths can be reduced stage by stage.
In the last stage, we just need to integrate a set of small and simple units (i.e., Dispatcher) to distribute access requests to consecutive output channels.

\subsection{MDP-network for Dataflow Propagation}
Referring to Fig. \ref{fig:001conflicts} \blackcircled{3}, the pattern of dataflow propagation is that multiple input data are directed to multiple output channels.
As MDP-network is designed to deal with such a pattern, we directly deploy original MDP-network to this stage to alleviate datapath conflicts and design centralization.

\section{Evaluation}

\subsection{Experimental Setup}

\begin{figure}[!t]
    \vspace{-10pt}
    \centering
    \includegraphics[page=5, width=0.28\textwidth]{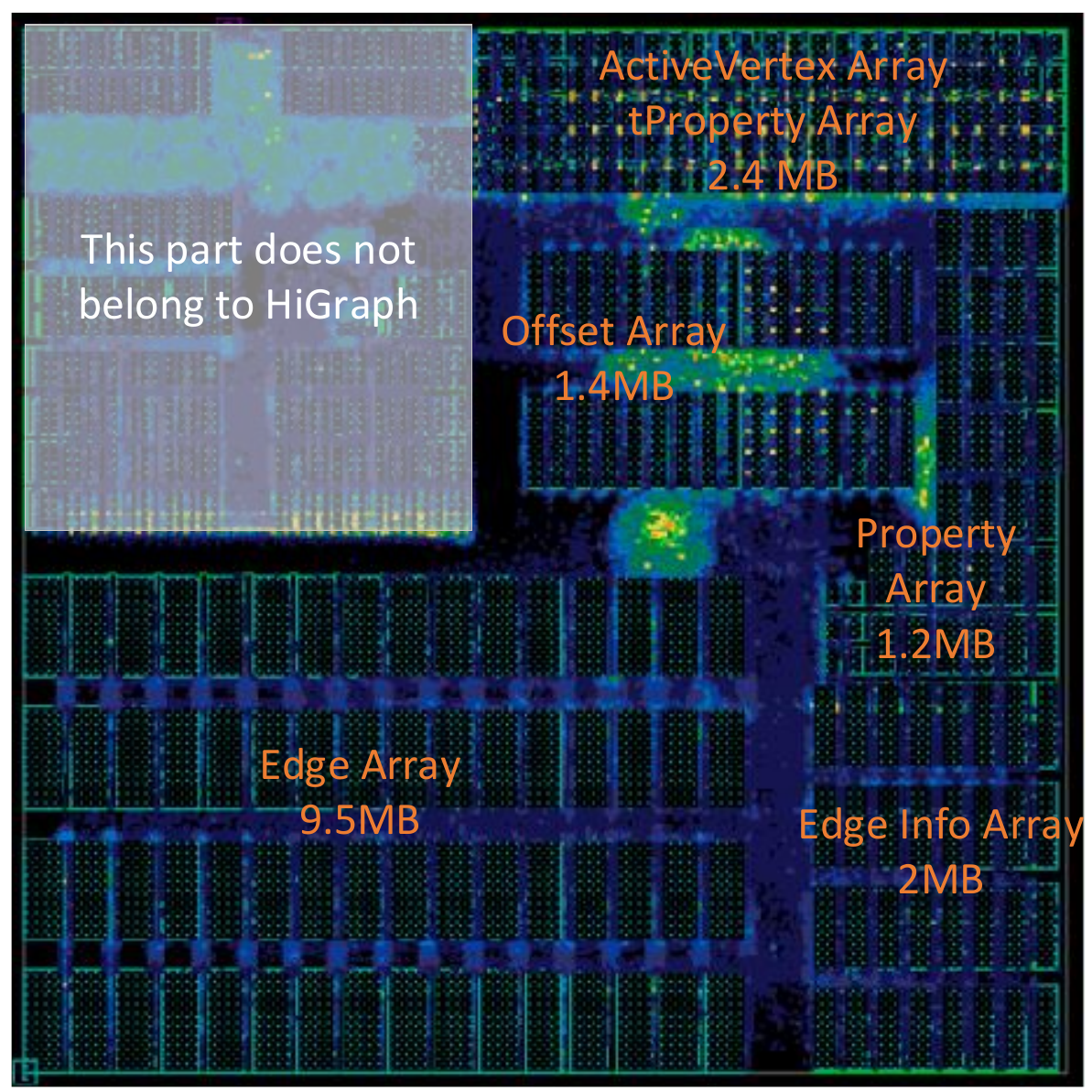}
    \vspace{-10pt}
    \caption{Layout of HiGraph.}
    \label{fig:005layout}
    \vspace{-5pt}
\end{figure}

\noindent\textbf{Methodology}.
We implement HiGraph in RTL with Verilog. We use the Synopsys Design Compiler with the TSMC 12$nm$ standard VT library for synthesis. We give the synthesis tools an operating voltage of 0.8$V$ and a target clock cycle of 1$ns$. The slowest module has a critical path of 0.93$ns$ including setup and hold time, putting the HiGraph design comfortably at 1GHz. The ID and property data of each vertex are quantified to 19 bits to fully use on-chip memory capacity. The layout of HiGraph is shown in Fig. \ref{fig:005layout}.

\begin{table}[!t]
 \vspace{-10pt}
 \caption{{Configurations used for HiGraph and baselines}.} \label{table:system}
 \vspace{-10pt}
 \centering
 \renewcommand\arraystretch{1.0}
    \resizebox{0.38\textwidth}{!}{
\begin{tabular}{|c|rrr|}
\hline
\textbf{}    & \textbf{HiGraph}    & \textbf{HiGraph-mini} & \textbf{GraphDynS} \\ \hline \hline
             Frequency             &1GHz       &1GHz   &1GHz   \\
			 \#Front-end channels  &32         &4      &4  \\
			 \#Back-end channels   &32         &32     &32  \\
			 On-chip memory        &16MB       &16MB   &32MB   \\ 
\hline 
\end{tabular}
    }
\end{table}

\noindent\textbf{Baselines}.
To compare HiGraph with state-of-the-art work, we prototype GraphDynS in RTL. 
We set the number of front-end channels in GraphDynS as four since a larger number would give rise to frequency decline due to the delicate arbitration in reading Offset Array. To compare fairly, we also set up HiGraph-mini with the same number of front-end channels. Table \ref{table:system} shows the configurations for these implementations.

\noindent\textbf{Datasets}.
Table \ref{table:dataset} describes the datasets used for our evaluation.
A mixture of real-world graphs - VT, EP, SL, TW and synthetic graphs - R14, R16 are used in the evaluation. Four graph algorithms - BFS (Breadth-First Search), SSSP (Single Source Shortest Path), SSWP (Single Source Widest Path) and PR (PageRank) are used to evaluate HiGraph. For the evaluation on unweighted graphs, random integer weights are assigned.

\begin{table}[!t]
 \vspace{-10pt}
 \caption{{Benchmark Datasets}.} \label{table:dataset}
 \vspace{-10pt}
 \centering
 \renewcommand\arraystretch{1.0}
    \resizebox{0.48\textwidth}{!}{
\begin{tabular}{|c|rrrr|}
\hline
\textbf{Name}    & \textbf{\#Vertices} & \textbf{\#Edges}  & \textbf{\#Degree} & \textbf{Description}  \\ \hline \hline
\multicolumn{5}{|c|}{\begin{tabular}[c]{@{}c@{}} \textbf{Real-world Graphs} \end{tabular}} \\ \hline
Vote (VT) \cite{DBLP:conf/www/LeskovecHK10/02VT}            & 7K       &0.10M    &15   &Wikepedia Who-votes-on-whom\\
Epinions (EP) \cite{DBLP:conf/semweb/RichardsonAD03/03EP}   & 76K      &0.51M    &7    &Epinions Who-trusts-whom\\
Slashdot (SL) \cite{DBLP:journals/im/LeskovecLDM09/04SL}    & 82K      &0.95M    &12   &Slashdot Social Network\\
Twitter (TW) \cite{DBLP:07TW}                               & 81K      &1.77M    &22   &Twitter Social Circles\\
\hline \hline
\multicolumn{5}{|c|}{\begin{tabular}[c]{@{}c@{}} \textbf{Synthetic Graphs} \end{tabular}} \\ \hline
RMAT14 (R14) \cite{07RMAT}    &16K &1.05M  &64 &Synthetic Graph\\
RMAT16 (R16) \cite{07RMAT}    &66K &4.19M  &64 &Synthetic Graph\\ \hline

\end{tabular}
    }
\end{table}

\subsection{Overall Results}

\begin{figure}[htbp]
    \vspace{-10pt}
    \centerline{\includegraphics[width=0.48\textwidth]{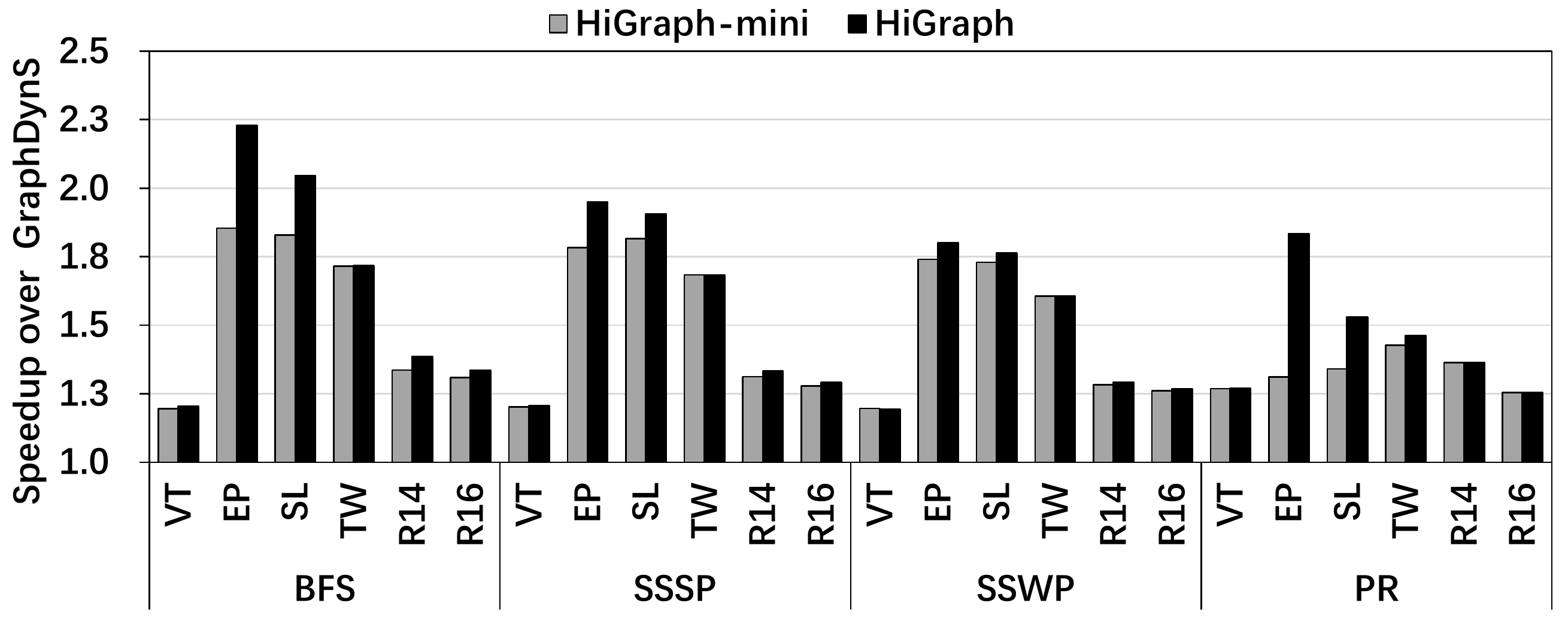}}
    \vspace{-10pt}
    \caption{Speedup over GraphDynS.}
    \label{fig:new02}
    \vspace{-5pt}
\end{figure}

\noindent\textbf{Speedup:} Fig. \ref{fig:new02} shows the speedups of HiGraph and HiGraph-mini normalized to GraphDynS. With the same number of front-end channels, HiGraph-mini achieves 1.19$\times$ to 1.85$\times$ speedup over GraphDynS, and 1.46$\times$ on average. With MDP-network optimization, HiGraph increases the number of front-end channels without frequency decline. 
Thus, with more front-end channels, HiGraph achieves up to 2.23$\times$ speedup over GraphDynS (1.54$\times$ on average).

\begin{figure}[htbp]
    \vspace{-7pt}
    \centerline{\includegraphics[width=0.48\textwidth]{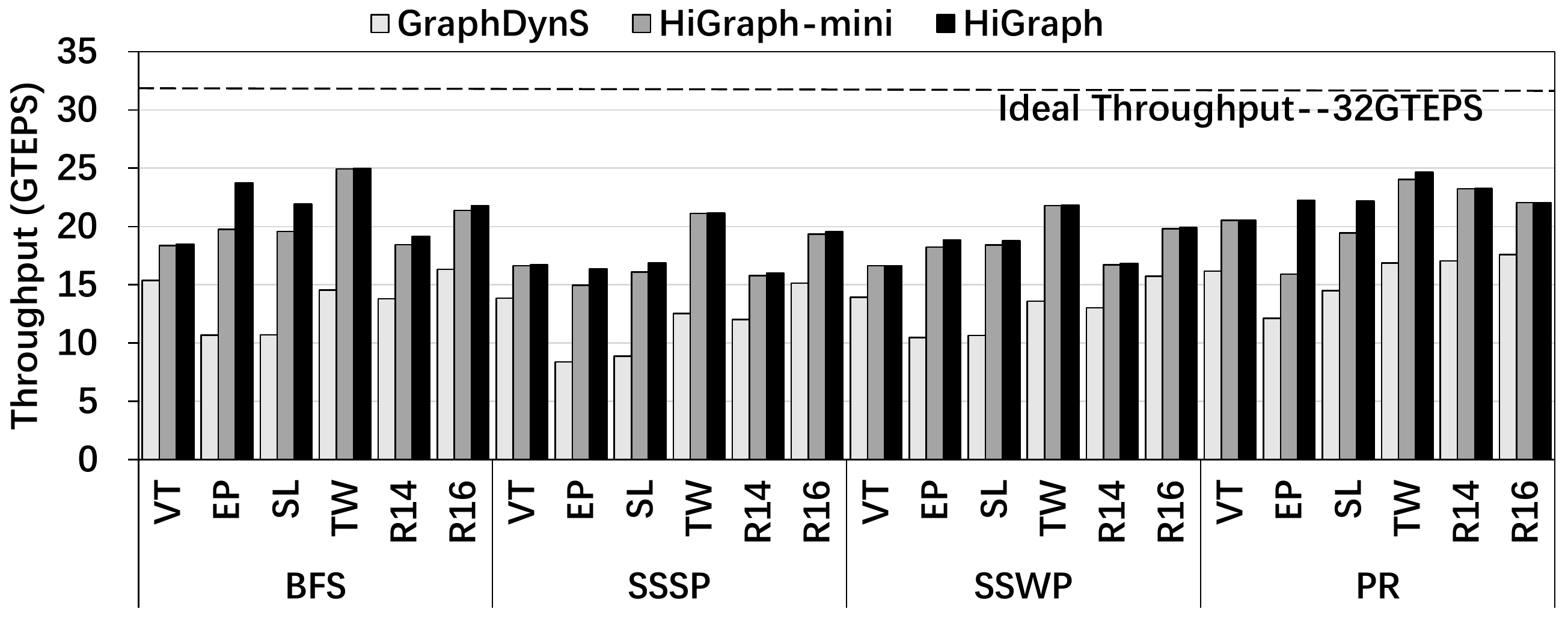}}
    \vspace{-10pt}
    \caption{Throughput.}
    \label{fig:new01}
    \vspace{-10pt}
\end{figure}

\noindent\textbf{Throughput:} Fig. \ref{fig:new01} compares the throughput of HiGraph against that of GraphDynS. Throughput is defined as the number of edges processed per second (GTEPS, giga-traversed edges per second). The ideal throughput is 32 GTEPS. HiGraph achieves up to 25.0 GTEPS and reaches 78.1\% of ideal throughput. Compared to GraphDynS, the throughput is improved by 2.7 GTEPS to 13.1 GTEPS, and 6.7 GTEPS on average.



\begin{figure*}[!htbp] 
    \centering
    \begin{minipage}{0.43\textwidth}
    \vspace{-2pt}
    \centering
    \includegraphics[width=\textwidth]{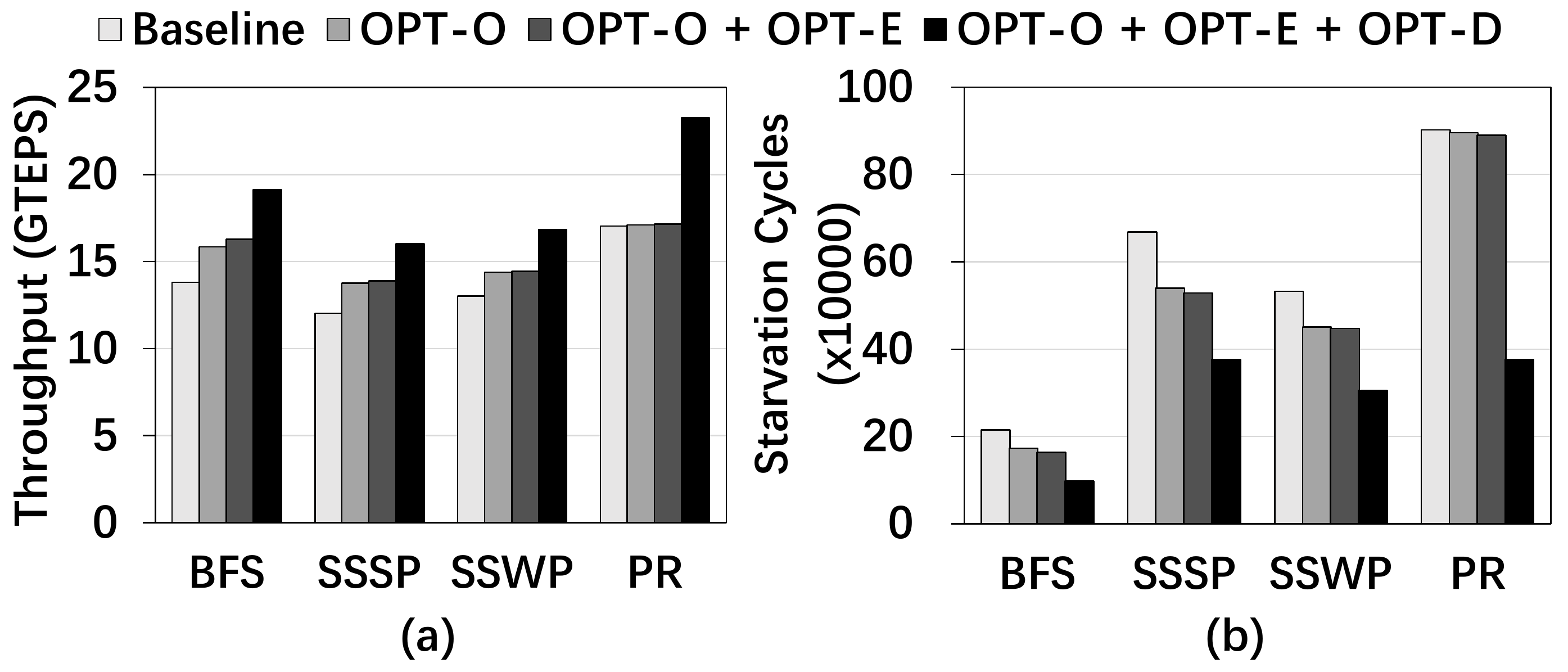}
    \vspace{-22pt}
    \caption{Effects of our optimizations: (a) Throughput. (b) Starvation cycles of vPE.}
    \label{fig:new04}
    \end{minipage}\hfill
    \begin{minipage}{0.22\textwidth}
    \centering
    \includegraphics[width=\textwidth]{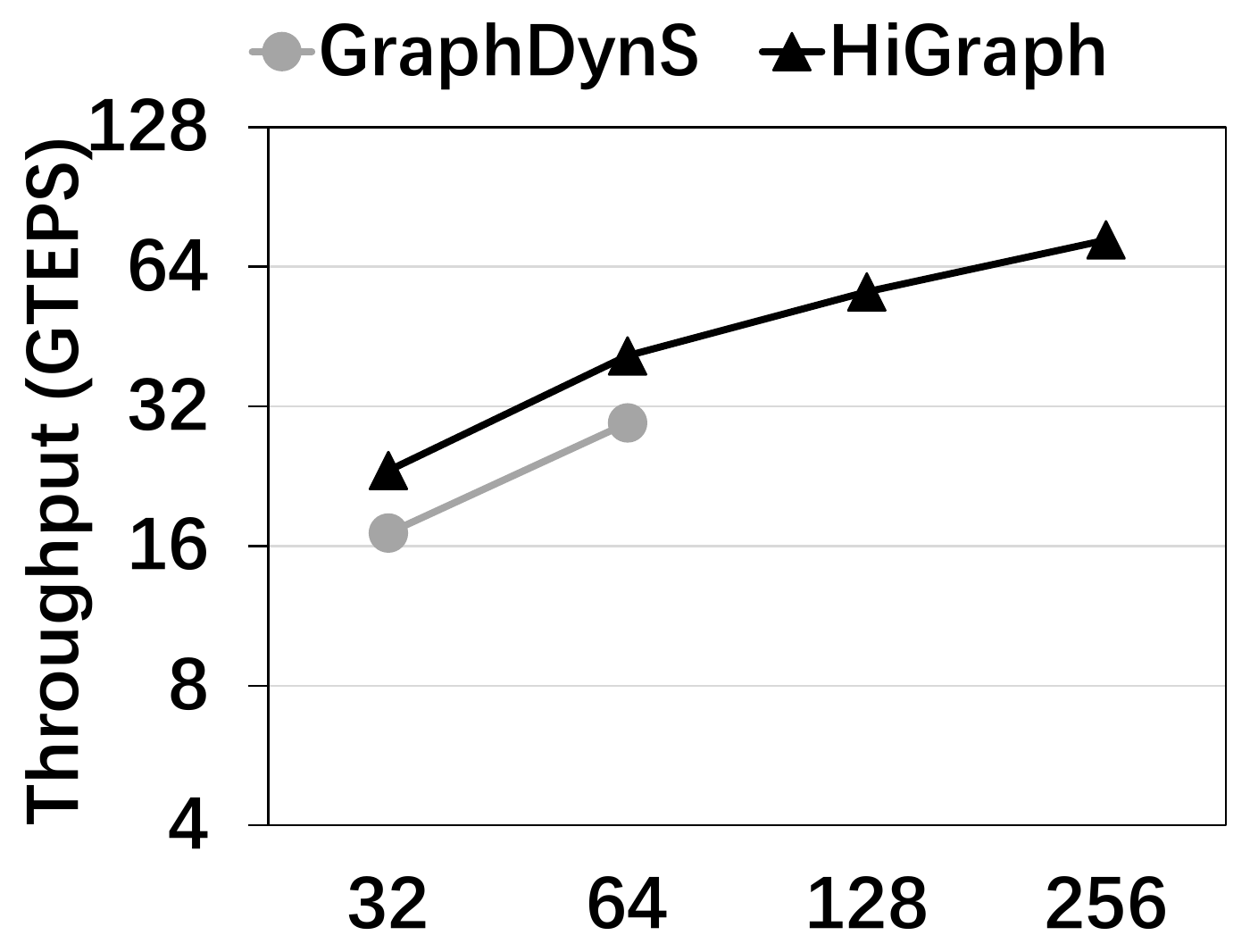}
    \vspace{-22pt}
    \caption{{Throughput versus number of back-end channels.}}
    \label{fig:new06}
    \end{minipage}\hfill
    \begin{minipage}{0.27\textwidth}
    \centering
    \includegraphics[width=\textwidth]{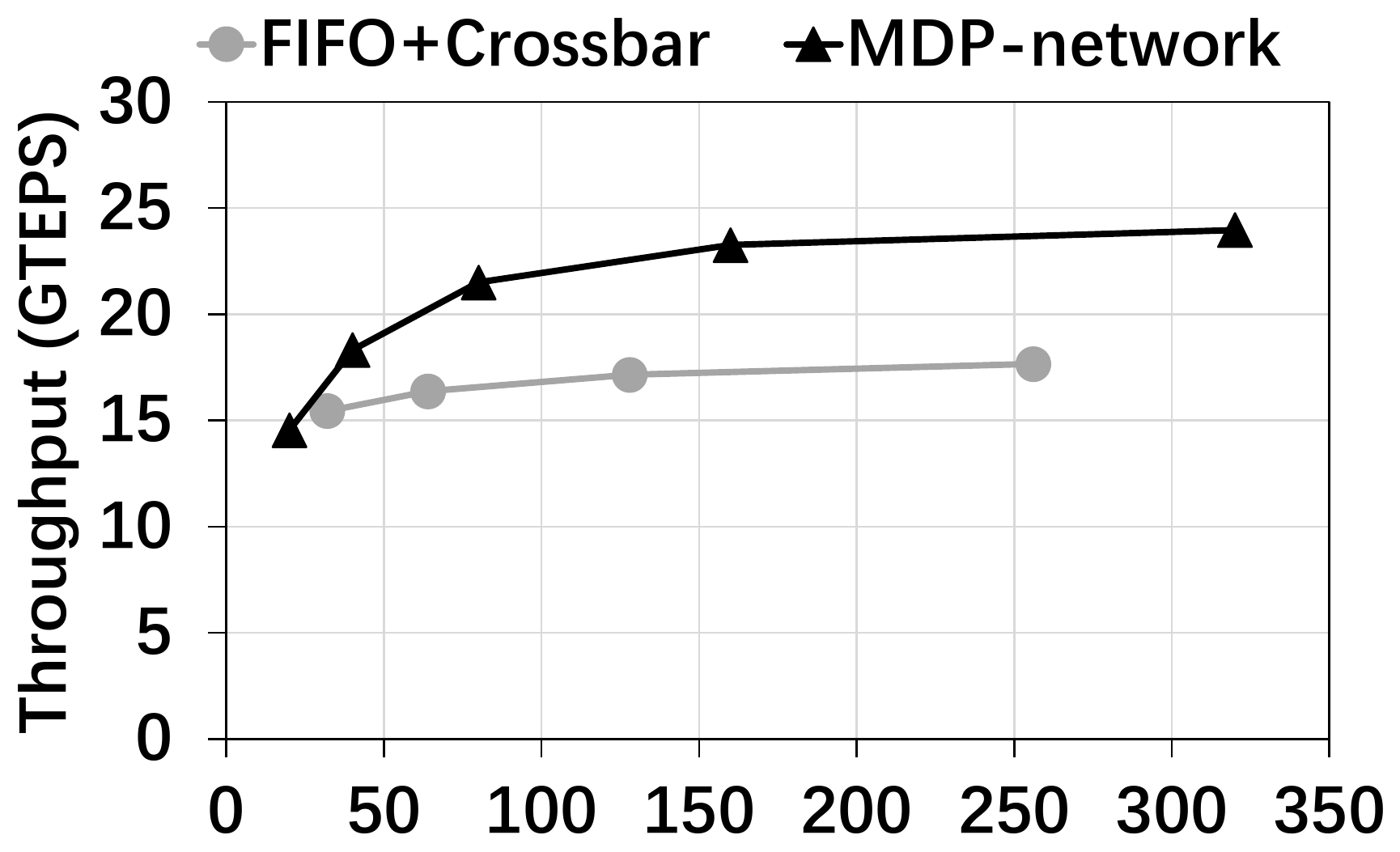}
    \vspace{-22pt}
    \caption{Throughput versus buffer size of FIFO in each channel.}
    \label{fig:new03}
    \end{minipage}\hfill
    \vspace{-15pt}
\end{figure*}

\subsection{Effects of Optimizations}
A detailed evaluation on the RMAT14 dataset is presented to provide more insights into the effects of our optimizations, as shown in Fig. \ref{fig:new04} (a). The baseline is without any our optimizations. Opt-O, Opt-E, and Opt-D indicate the use of MDP-network for Offset Array Access, Edge Array Access, and Dataflow Propagation, respectively.  
Following two obvious phenomenons can be found in Fig. \ref{fig:new04} (a).
First, when using Opt-D in optimization, the design gains more performance improvement, up to 6.2 GTEPS.
This is because Opt-D not only alleviates the datapath conflict in back-end part, but also has synergy with the optimizations in front-end part.
Second, the optimizations in front-end part almost gain no performance improvement on the PR algorithm. 
This is because the Offset Array and Edge Array are read in order on the PR algorithm, so that no datapath conflict arises in frond-end part.

To show the reduction of datapath conflicts, we provide the number of starvation cycles of vPE to reveal that. The vPE performs the Reduce( ) function, the final step in the scatter phase. Fig. \ref{fig:new04} (b) shows a massive number of starvation cycles of baseline on the RMAT14 dataset, which reveals that plenty of vPEs are starved for data due to the datapath conflicts. 
Benefiting from our optimizations, datapath conflicts are alleviated so that more data is transferred to vPE without stall. Thus, the number of starvation cycles reduces significantly, up to 58\%.
This validates the effects of our optimizations.


Fig. \ref{fig:new06} shows the scalability of HiGraph and GraphDynS on the PR algorithm with RMAT14 dataset as the number of back-end channels increases. 
GraphDynS does not support more than 64 channels due to significant frequency decline as shown in Fig. \ref{fig:002freq_cb}, so we only evaluate GraphDynS with 32 and 64 channels.
For HiGraph, we synthesis it from 32 to 256 channels using Synopsys Design Compiler and find that the most critical path of it only raising from 0.93ns to 0.97ns, still meeting the requirement of 1GHz. The result in Fig. \ref{fig:new06} demonstrates that HiGraph's scalability is much better than GraphDynS.



\textbf{Discussion:} For the large graph processing, the graph can be partitioned into small slices, so that each slice is processed on chip \cite{Graphicionado}. Therefore, our optimizations can improve throughput in large-scale graph analytics. Besides, the time consumed in the replacement of slices can be overlapped using double buffer design.

\subsection{Design Option of MDP-network}

A detailed evaluation of design option is presented to provide more insights into the design of MDP-network.

We run experiments on various radices, i.e., the write port number of the FIFO used to construct the stage of MDP-network. 
We find that a too large radix still encounters design centralization, which degrades the performance. By contrast, the performance changes slightly with relatively small radices. Thus, we choose radix 2 in our design.


We run experiments on buffer size of MDP-network.
We keep all designs in HiGraph the same except for the dataflow propagation stage, in which we replace MDP-network with FIFO-plus-crossbar design. Fig. \ref{fig:new03} demonstrates that MDP-network outperforms FIFO-plus-crossbar consistently with various buffer sizes on the PR algorithm with RMAT14 dataset. We choose 160 entries as the buffer size of FIFO in each channel because the throughput rarely increases with larger buffers. We synthesis MDP-network with buffer size 160 and the area is 0.375$mm^2$ while power is 621.2mW. We also synthesis FIFO-plus-crossbar design with buffer size 128 and the area is 0.292$mm^2$ while power is 508.1mW. The area and power of MDP-network is slightly higher due to the larger buffer, showing that replacing crossbar with MDP-network brings little overhead.



\section{Related Work}
Prior works focus on optimizing irregular off-chip memory accesses to pursue high throughput of graph analytics~\cite{Graphicionado,GraphDynS, GraphPulse, ASIC_ISCA2016_multi}. 
By preprocessing \cite{GraphChi} and partitioning \cite{Graphicionado}, a large graph can be partitioned into a set of slices to fit the data being irregularly accessed in on-chip memory. Moreover, Centaur \cite{dac:centaur} is an accelerator that only maps high-degree vertices to on-chip memory. To further pursue higher throughput, prior accelerators \cite{Graphicionado, GraphDynS} provide parallel execution channels design according to the execution characteristic of graph analytics workloads.
Unfortunately, few attention is paid to datapath conflicts and design centralization that have become the critical issues. 
Using our MDP-network, these issues are alleviated and high throughput is realized.

\section{Conclusion}

In this paper, we identify the inefficiencies in graph analytics acceleration including the datapath conflicts and design centralization. 
To this end, we propose MDP-network, inspired by the idea of trading latency for throughput. Besides, an automatic generator for MDP-network is developed and open source.
Finally, a novel high throughput graph analytics accelerator, HiGraph, is proposed by deploying MDP-network to tackle data conflicts and design centralization in practice. HiGraph archives up to 2.2$\times$ speedup (1.5$\times$ on average) compared to state-of-the-art accelerator.



\begin{acks}
This work was supported by the National Natural Science Foundation of China (Grant No. 61732018, 61872335, and 61802367), Austrian-Chinese Cooperative R\&D Project (FFG and CAS) (Grant No. 171111KYSB20200002), CAS Project for Young Scientists in Basic Research (Grant No. YSBR-029), and CAS Project for Youth Innovation Promotion Association. The correponding author is Mingyu Yan, yanmingyu@ict.ac.cn.
\end{acks}

\bibliographystyle{ACM-Reference-Format}
\bibliography{higraph_arxiv}

\end{document}